\documentclass[aps,pre,reprint,groupedaddress,showpacs]{revtex4-1}
\usepackage[dvipdfmx]{graphicx}
\usepackage{latexsym}
\usepackage[fleqn]{amsmath}
\usepackage{amssymb}
\usepackage{bm}

\usepackage{amsthm}
\usepackage{mathrsfs}  
\usepackage{dcolumn}

\begin{document}

\newtheorem{Def}{Definition}[section]
\newtheorem{Thm}{Theorem}[section]
\newtheorem{Proposition}{Proposition}[section]
\newtheorem{Lemma}{Lemma}[section]
\theoremstyle{definition}
\newtheorem*{Proof}{Proof}%[section]
\newtheorem{Example}{Example}[section]
\newtheorem{Postulate}{Postulate}[section]
\newtheorem{Corollary}{Corollary}[section]
\newtheorem{Remark}{Remark}[section]
\theoremstyle{remark}
\newcommand{\beq}{\begin{equation}}
\newcommand{\beqa}{\begin{eqnarray}}
\newcommand{\eeq}{\end{equation}}
\newcommand{\eeqa}{\end{eqnarray}}
\newcommand{\non}{\nonumber}
\newcommand{\lb}{\label}
\newcommand{\fr}[1]{(\ref{#1})}
\newcommand{\cc}{\mbox{c.c.}}
\newcommand{\nr}{\mbox{n.r.}}
\newcommand{\bb}{\mbox{\boldmath {$b$}}}
\newcommand{\bbe}{\mbox{\boldmath {$e$}}}
\newcommand{\bt}{\mbox{\boldmath {$t$}}}
\newcommand{\bn}{\mbox{\boldmath {$n$}}}
\newcommand{\br}{\mbox{\boldmath {$r$}}}
\newcommand{\bC}{\mbox{\boldmath {$C$}}}
\newcommand{\bH}{\mbox{\boldmath {$H$}}}
\newcommand{\bp}{\mbox{\boldmath {$p$}}}
\newcommand{\bx}{\mbox{\boldmath {$x$}}}
\newcommand{\bF}{\mbox{\boldmath {$F$}}}
\newcommand{\bT}{\mbox{\boldmath {$T$}}}
\newcommand{\bomega}{\mbox{\boldmath {$\omega$}}}
\newcommand{\ve}{{\varepsilon}}
\newcommand{\e}{\mathrm{e}}
\newcommand{\F}{\mathrm{F}}
\newcommand{\Loc}{\mathrm{Loc}}
\newcommand{\hF}{\widehat F}
\newcommand{\hL}{\widehat L}
\newcommand{\tA}{\widetilde A}
\newcommand{\tB}{\widetilde B}
\newcommand{\tC}{\widetilde C}
\newcommand{\tL}{\widetilde L}
\newcommand{\tK}{\widetilde K}
\newcommand{\tX}{\widetilde X}
\newcommand{\tY}{\widetilde Y}
\newcommand{\tU}{\widetilde U}
\newcommand{\tZ}{\widetilde Z}
\newcommand{\talpha}{\widetilde \alpha}
\newcommand{\te}{\widetilde e}
\newcommand{\tv}{\widetilde v}
\newcommand{\ts}{\widetilde s}
\newcommand{\tx}{\widetilde x}
\newcommand{\ty}{\widetilde y}
\newcommand{\ud}{\underline{\delta}}
\newcommand{\uD}{\underline{\Delta}}
\newcommand{\chN}{\check{N}}
\newcommand{\cA}{{\cal A}}
\newcommand{\cB}{{\cal B}}
\newcommand{\cC}{{\cal C}}
\newcommand{\cD}{{\cal D}}
\newcommand{\cF}{{\cal F}}
\newcommand{\cL}{{\cal L}}
\newcommand{\cM}{{\cal M}}
\newcommand{\cR}{{\cal R}}
\newcommand{\cS}{{\cal S}}
\newcommand{\cY}{{\cal Y}}
\newcommand{\cU}{{\cal U}}
\newcommand{\cV}{{\cal V}}
\newcommand{\tcA}{\widetilde{\cal A}}
\newcommand{\DD}{{\cal D}}
\newcommand\TYPE[3]{ \underset {(#1)}{\overset{{#3}}{#2}}  }
\newcommand{\bfe}{\boldsymbol e} 
\newcommand{\bfb}{{\boldsymbol b}}
\newcommand{\bfd}{{\boldsymbol d}}
\newcommand{\bfh}{{\boldsymbol h}}
\newcommand{\bfj}{{\boldsymbol j}}
\newcommand{\bfn}{{\boldsymbol n}}
\newcommand{\bfA}{{\boldsymbol A}}
\newcommand{\bfB}{{\boldsymbol B}}
\newcommand{\bfJ}{{\boldsymbol J}}
\newcommand{\dr}{\mathrm{d}}
\newcommand{\TE}{\mathrm{TE}}
\newcommand{\TM}{\mathrm{TM}}
\newcommand{\Ai}{\mathrm{Ai}}
\newcommand{\Bi}{\mathrm{Bi}}
\newcommand{\sech}{\mathrm{sech}}
\newcommand{\jthree}{  \TYPE 3  {j}  {}   }
\newcommand{\Lam}{ \TYPE q  {\Lambda}   {}   }
\newcommand{\alp}{ \TYPE p  {\alpha}   {}   }
\newcommand{\al}[1]{ \TYPE {#1}  {\alpha}   {}   }
\newcommand{\bep}{ \TYPE p  {\beta}   {}   }
\newcommand{\be}[1]{ \TYPE {#1}  {\beta}   {}   }
\newcommand{\gamq}{ \TYPE q  {\gamma}   {}   }
\newcommand{\hash}{\#}
\newcommand{\hashat}{\widehat{\#}}
\newcommand{\hashch}{\stackrel{\vee}{\#}}
\newcommand{\chd}{\stackrel{\vee}{\D}}
\newcommand\NN[1]{{\cal N}_{#1}}
\newcommand\MM[1]{{\cal M}_{#1}}
\newcommand\BAE[1]{{\begin{equation}{\begin{aligned}#1\end{aligned}}\end{equation}}}
\newcommand{\GamCLamM}[1]{{\Gamma\mathbb{C}\Lambda^{{#1}}\cal{M}}}
\newcommand{\GamLamM}[1]{{\Gamma\Lambda^{{#1}}\,\cal{M}}}
\newcommand{\GamLamU}[1]{{\Gamma\Lambda^{{#1}}\,\cal{U}}}
\newcommand{\GamLamHU}[1]{{\Gamma\Lambda^{{#1}}\,\widehat{\cal{U}}}}
\newcommand{\GamLam}[2]{{\Gamma\Lambda^{{#1}}\,{#2}}}
\newcommand{\GTM}{{\Gamma T\cal{M}}}
\newcommand{\GTU}{{\Gamma T\cal{U}}}
\newcommand{\GT}[1]{{\Gamma T {#1}}}
\newcommand{\normM}[2]{\left(  #1\, , \, #2 \right)}
\newcommand{\normU}[2]{\left\{ #1\, , \, #2 \right\}}
\newcommand{\diag}[1]{\mbox{diag}\{\, #1\,\}}
\newcommand{\GtM}[2]{\Gamma T^{#1}_{#2}{\cal M}}
\newcommand{\inp}[2]{\left\langle\,  #1\, , \, #2\, \right\rangle}
\newcommand{\defi}{\noindent {\bf Definition : } }
\newcommand{\prop}{\noindent {\bf Proposition : } }
\newcommand{\theo}{\noindent {\bf Theorem : } }
\newcommand{\exam}{\noindent {\bf Example : } }
\newcommand{\equp}[1]{\overset{\mathrm{#1}}{=}}
\newcommand{\wt}[1]{\widetilde{#1}}
\newcommand{\wh}[1]{\widehat{#1}}
\newcommand{\ch}[1]{\check{#1}}
\newcommand{\ii}{\imath}
\newcommand{\ic}{\iota}
\newcommand{\mi}{\,\mathrm{i}\,}
\newcommand{\mr}{\,\mathrm{r}\,}
\newcommand{\mbbC}{\mathbb{C}}
\newcommand{\mbbR}{\mathbb{R}}
\newcommand{\mbbZ}{\mathbb{Z}}
\newcommand{\Leftrightup}[1]{\overset{\mathrm{#1}}{\Longleftrightarrow}}
%\newcommand{\ol}[1]{\overline{#1}}
%\preprint{}

\title{Conditional Lyapunov Exponent Criteria in terms of Ergodic Theory}


\author{Masaru Shintani}
\email[shintani.masaru.28a@kyoto-u.jp]{}%{Your e-mail address}
\author{Ken Umeno}%
\email[umeno.ken.8z@kyoto-u.ac.jp]{}%{Your e-mail address}
\affiliation{Department of Applied Mathematics and Physics, Graduate
School of Informatics, Kyoto University, Yoshida Honmachi Sakyo-ku,
Kyoto 606--8501, Japan}


\date{\today}

 \begin{abstract}
The conditional Lyapunov exponent is defined for investigating chaotic
  synchronization, in particular complete synchronization and generalized
  synchronization.
  We find that the conditional Lyapunov exponent is expressed as a
  formula in terms of ergodic theory. %when dynamical system has ergodicity.
  Dealing with this formula, we find what factors characterize the
  conditional Lyapunov exponent in chaotic systems.
 \end{abstract}
%02.60.Cb, 
%\keywords{02.60.Cb, }

\maketitle

 \section{Introduction}
The conditional Lyapunov exponent is defined for investigating chaotic
synchronization \cite{fujisaka1983stability,pecora1990synchronization,kocarev1995general,kocarev1996generalized,yu1990transition,pikovsky1997coherence,pikovsky1997phase,pikovsky2003synchronization,boccaletti1999synchronization,boccaletti2002synchronization,sunghwan2000chaotic,guan2006understanding,lai1998synchronization,balanov2008synchronization},
in particular Complete synchronization (CS) and
Generalized synchronization (GS).
Transitions from desynchronization to synchronization of trajectories occur when the
conditional Lyapunov exponent changes from positive to negative \cite{goto2008conditional,sunada2014optical,uchida2006consistency}.
Although it is widely known that the chaotic synchronization %actually
occurs in many systems,
it is not clearly known %the mechanism of such synchronization.
why the conditional Lyapunov exponent changes.
For example, it has not completely been clarified why the CS occurs in
chaotic systems.
A report \cite{boccaletti2002synchronization} showed that an external forcing input in CS may change the dynamical system to another
one. The forcing input changes the balance between phase contraction or
expansion, and the CS occurs when such contraction dominates.
Although the explanation is well considered,
there could be another reason why the
conditional Lyapunov exponent may change.
Furthermore, they have focused on the mean of external forcing inputs
for the CS. The study \cite{sunghwan2000chaotic} showed that they got the transversal
Lyapunov exponent with changing the noise distribution.
These may be important. However we expect that more precise
information on phase space can lead to more relevant analysis.
Therefore, we would like to clarify
the relation between the conditional Lyapunov exponent and the chaotic
synchronization in CS.

We have two claims through this research.
The first one is about a formula to express the conditional Lyapunov exponent in
terms of ergodic theory,
and the second is what factors influence the conditional Lyapunov exponent in
chaotic dynamical systems.
Although existing research offered some mechanisms of
chaotic synchronizations by focusing on  mean
amplitudes or variances of common input signals ( See \cite{guan2006understanding,lai1998synchronization} for example),
we find that such mean amplitudes or variances of common input signals
are not imperative. Instead, we reveal that
a {\it distribution characteristic} of input signals is the
most important factor. We also reveal that
this characteristic determines transitions between
synchronization and desynchronization, and this does not
depend on whether input signals are chaotic or noisy.
Thus, although the second claim can easily be
derived from the first one, we emphasize that our second claim is
physically important.

In Section 2, we describe the definition of conditional Lyapunov exponent in chaotic
systems. Furthermore, we describe two main claims in our research.

In Sections 3 - 5, we construct solvable chaotic dynamical systems, and
confirm the our claims by analysis for such systems.

In Appendix A, we summarize a short introduction of ergodic theory.

\section{Definition and Main Claim}
To discuss the conditional Lyapunov exponent,
we consider the following simple one-dimensional unidirectional coupling
system:
\begin{eqnarray}\label{eq:system1}
  x(t+1)=f(x(t))+\xi(t)\equiv \psi(x(t),\xi(t)),
\end{eqnarray}
where $t$ is time, $x(t)$ is a response,
$f$ is a chaotic mapping,
 and $\xi(t)$ is an external forcing driver.
We define ${\rm P}_{X}(X)$ as the probability distribution of variables
 $X$ for the dynamical system, and also define
 $\mathbb{R}_{X}$ as the range in which variables $X$ are defined.

We define the CS of the system as follows
\cite{pikovsky2003synchronization,boccaletti1999synchronization,boccaletti2002synchronization,sunghwan2000chaotic,goto2008conditional}.
We consider two different trajectories $x_1(t)$ and $x_2(t)$ in
Eq. \eqref{eq:system1} with different initial points.
To judge whether the response system exhibit CS, we introduce the %laregest
conditional Lyapunov exponent.
The occurrence of CS implies that the difference between $x_1(t)$ and
$x_2(t)$ decreases as time increases.
The CS is said to occur when the following condition about synchronization error $e'(t)\equiv |x_2(t)-x_1(t)|$:
\begin{eqnarray}\label{eq:synchronization}
\lim_{t\to \infty}e'(t)=0,
\end{eqnarray}
is satisfied for almost all initial points. Here $|...|$ expresses Euclidean norm of
the argument.
The interpretation of \eqref{eq:synchronization} is that the state in
the large $t$ limit is independent of any initial state for almost all
initial points.
The infinitesimal synchronization error to the projection to $x$ axis $e(T)$ in system
\eqref{eq:system1} is defined as:
\begin{eqnarray}\label{eq:delta}
 e(T):=\prod_{t=0}^{T}\left|\frac{\partial \psi(x,\xi)}{\partial x}\right|_{x=x(t)}e(0).
\end{eqnarray}
When CS occurs, the equation $\lim_{t\to \infty}e'(t)=\lim_{T\to
\infty}|e(T)|=0$ is satisfied.
It is noted here that $\xi(t)$ in Eq. \eqref{eq:system1} affects the
time-evolution of $x(t)$ and contributes to $e(T)$.

The conditional Lyapunov exponent for variables $x$ is
defined by
\begin{eqnarray*}
\lambda:=\lim_{T \to \infty}\frac{1}{T}\ln\frac{|e(T)|}{|e(0)|}.
\end{eqnarray*}
Clearly $\lambda$ expresses the stability of the state
$x_1(t)=x_2(t)$. In this paper, the CS is said to occur when
$\lambda<0$, and we do not use Eq. \eqref{eq:synchronization} directly. 
The conditional Lyapunov exponent exists 
when the system \eqref{eq:system1} is ergodic with respect to $x$ and $\xi$
according to Oseledets' multiplicative ergodic theorem for autonomous dynamical systems \cite{oseledets1968multiplicative}.

Here, we assume that the system \eqref{eq:system1} can be seen as a two-dimensional autonomous dynamical system of $x$ and $\xi$.
When the system \eqref{eq:system1} is ergodic with respect to the absolutely continuous invariant measure (physical measure) with respect to $x$ and $\xi$, the conditional Lyapunov exponent $\lambda$ is
expressed as the ensemble average:
 \begin{eqnarray}\label{eq:pastdif}
  \lambda&=&\lim_{T \to
   \infty}\frac{1}{T}\ln\frac{|e(T)|}{|e(0)|}\notag\\
    &=& \int_{\mathbb{R}_{x}}\int_{\mathbb{R}_{\xi}}{\rm
     P}(x,\xi)\ln\left|\frac{\partial \psi(x,\xi)}{\partial
			   x}\right|{\rm d}\xi{\rm d}x,
\end{eqnarray}
where, ${\rm P}(x,\xi)$ is the invariant distribution of the system \eqref{eq:system1}.

We can generalize Eq. \eqref{eq:pastdif} to higher dimensional dynamical systems. 
We define the following two-dimensional dynamical system with an external forcing input:
\begin{eqnarray}\label{eq:two-dimensionnal}
\left\{
\begin{aligned}
x_1(t+1) &= f_1(x_1(t),x_2(t))+\xi(t) \\
&\equiv \psi_1(x_1(t),x_2(t),\xi(t))\\
x_2(t+1) &= f_2(x_1(t),x_2(t))+\xi(t) \\
&\equiv \psi_2(x_1(t),x_2(t),\xi(t)).
\end{aligned}
\right.
\end{eqnarray}
If the two-dimensional dynamical system \eqref{eq:two-dimensionnal} is ergodic with respect to the absolutely continuous invariant measure in terms of $x_1$ and $x_2$ variables, the conditional Lyapunov exponent $\lambda_{ij} (i=1,2, j=1,2)$, which is defined by the infinitesimal synchronization error for $\psi_{j}$ to the projection to $x_{i}$ axis,
is defined by

\begin{eqnarray*}
\lambda_{ij} &=&\lim_{T\to\infty}\frac{1}{T}\ln\frac{\left|e_{\psi_{j}}(T)\right|}{\left|e_{x_{i}}(0)\right|}\\
 &=&\int_{\mathbb{R}_{x_{1}}}\int_{\mathbb{R}_{x_{2}}}\int_{\mathbb{R}_\xi}{\rm P}(x_{1},x_{2},\xi)\ln\left|\frac{\partial\psi_{j}(x_{1},x_{2},\xi)}{\partial x_{i}}\right|{\rm d}\xi{\rm d}x_1{\rm d}x_2.
\end{eqnarray*}

In what follows, however we assume that the system \eqref{eq:system1} has the one-dimensional limiting distribution given by
the absolutely continuous invariant ergodic measure
$\mu({\rm d}x)={\rm P}_{x}(x){\rm d}x,(x \in \mathbb{R}_x)$ with some ${\rm P}_x$, and ${\rm P}(x,\xi)$ is a continuous density function in $x$ and $\xi$ satisfying $\int_{\mathbb{R}_x}\int_{\mathbb{R}_\xi} {\rm P}(x,\xi){\rm d}\xi{\rm d}x=1$ for simplicity.

Then, we give our two main claims.
 Firstly, the conditional Lyapunov exponent is expressed by the following
 equation.
\begin{eqnarray}
 \lambda= \tilde{\lambda} \label{eq:conclusion1},
\end{eqnarray}
where
\begin{eqnarray}
\lambda &=& \int_{\mathbb{R}_{x}}\int_{\mathbb{R}_{\xi}}{\rm
     P}(x,\xi)\ln\left|\frac{\partial \psi(x,\xi)}{\partial
		  x}\right|{\rm d}\xi{\rm d}x\label{eq:conventional}\\
 \tilde{\lambda} &=& \int_{\mathbb{R}_{\xi}}{\rm P}_{\xi}(\xi_0)\lambda_{\Psi}(\xi_0){\rm d}\xi_0\label{eq:proposing}.
\end{eqnarray}
Note that although $\xi_0$ is originally a constant value, we consider the {\it distribution} of $\xi_0$ in the above equation.
 Here $\lambda_\Psi(\xi_0)$ is defined as follows.
 First, by calculating the Lyapunov exponent $\lambda_\Psi(\xi_0)$ for the
 auxiliary system:
\begin{eqnarray}\label{eq:Psi}
y(t+1)=\Psi_{\xi_0}(y(t))
\end{eqnarray}
with $\Psi_{\xi_0}(y)=f(y)+\xi_0$.
Then, by changing $\xi_0$ continuously, we have the set
$\{\lambda_\Psi(\xi_0)|\xi_0\in \mathbb{R}_\xi\}$.
Note that the functions $\Psi_{\xi_0}$ are one-dimensional functions with the parameters $\xi_0$ distributed according to ${\rm P}(\xi)$, while $\psi(y,\xi_0)$ is just a two-dimensional function which is different from $\Psi_{\xi_0}(y)$.

Equation \eqref{eq:conclusion1} implies that the conditional Lyapunov
exponent $\lambda$ is expressed as
{\it the ensemble average of the set of unique Lyapunov exponents}
$\{\lambda_\Psi(\xi_0)|\xi_0\in \mathbb{R}_\xi\}$
provided the ergodicity in the one-dimensional dynamical system \eqref{eq:Psi}.
This is the first main claim in our research.

\begin{figure}[htb]
  \begin{center}
     \includegraphics[width=80mm]{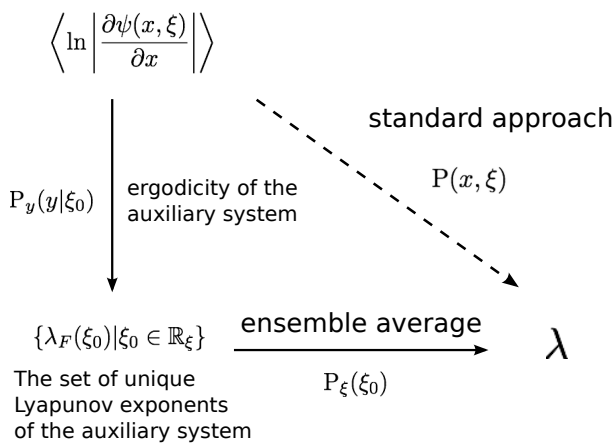}
    \begin{center}
     \caption{Our approach for the conditional Lyapunov exponent}
      \label{fig:new}
    \end{center}
  \end{center}
\end{figure}

Then, we show how we get this claim provided the
ergodicity in the system \eqref{eq:system1}.
The Lyapunov exponent $\lambda_{\Psi}(\xi_0)$ can be obtained by the
following equation with the ergodicity in the system \eqref{eq:Psi},
\begin{eqnarray}\label{eq:lambda_Psi}
 \lambda_\Psi(\xi_0) &=& \int_{\mathbb{R}_{y}} {\rm P}_{y}(y|\xi_0)\ln\left|\frac{{\rm d} \Psi_{\xi_0}(y)}{{\rm d} y}\right|{\rm d}y,
\end{eqnarray}
where ${\rm P}(y|\xi_0)$ is the unique {\it conditional
probability distribution}
in dynamical system \eqref{eq:Psi} with each constant value $\xi_0$.
To obtain Eq. \eqref{eq:proposing}
we use the {\it marginalization} about random variables $a$ and $b$ in
the probability theory:
\begin{eqnarray}\label{eq:marginalization}
  {\rm P}(a,b)={\rm P}(b){\rm P}(a|b).
\end{eqnarray}
Then, substituting \eqref{eq:marginalization} into \eqref{eq:conventional}, we have
\begin{eqnarray}
\lambda &=& \int_{\mathbb{R}_{y}}\int_{\mathbb{R}_{\xi}}{\rm P}_{\xi}(\xi_0) {\rm
     P}_{y}(y|\xi_0)\ln\left|\frac{{\rm d}\Psi_{\xi_0}(y)}{{\rm d}
			   y}\right|{\rm d}\xi_0{\rm d}y\notag \\
 &=& \int_{\mathbb{R}_{\xi}} {\rm
  P}_{\xi}(\xi_0)\left(\int_{\mathbb{R}_{y}} {\rm
		P}_{y}(y|\xi_0)\ln\left|\frac{{\rm d}\Psi_{\xi_0}(y)}{{\rm d}y}\right|{\rm d}y\right){\rm d}\xi_0\notag\\
 &=&\int_{\mathbb{R}_{\xi}}{\rm P}_{\xi}(\xi_0)\lambda_{\Psi}(\xi_0){\rm
  d}\xi_0\label{eq:CLE_solution} =\tilde{\lambda}.
\end{eqnarray}
Note that $\lambda=\tilde{\lambda}$ is satisfied when the invariant distribution
${\rm P}_{y}(y|\xi)$ exists.
This is satisfied when the system \eqref{eq:Psi} is ergodic for any $\xi_0$.
Therefore, our first claim Eq. \eqref{eq:conclusion1} is derived from
the mixing property of \eqref{eq:system1} and the ergodicity of $\xi$
and the system \eqref{eq:Psi}.
Figure \ref{fig:new} illustrates this concept. Here, $\langle
... \rangle$ expresses the ensemble average with respect to the ergodic
invariant measure.
Note that if we have two-dimensional dynamical systems, the conditional Lyapunov exponent $\lambda_{ij} (i=1,2, j=1,2)$, which is defined by the infinitesimal synchronization error to the projection to $x_i$ axis, is expressed as:
\begin{eqnarray*}
\lambda_{ij} 
=
&\int_{\mathbb{R}_{x_1}}\int_{\mathbb{R}_{x_2}}\int_{\mathbb{R}_\xi}{\rm P}_{\xi}(\xi_0){\rm P}_{x_1,x_2}(x_1,x_2|\xi_0)\\
&\ln\left|\frac{\partial{\Psi_j}_{\xi_0}(x_1,x_2)}{\partial x_i}\right|{\rm d}\xi_0{\rm d}x_1{\rm d}x_2.\\
\end{eqnarray*}

As for the second claim, with the first one, we can derive that
the conditional Lyapunov exponent in a system
\eqref{eq:system1} is characterized by only
two factors, a dynamical system and a distribution of external forcing input.

\section{Example1 for the first claim}
In order to confirm the first claim, we would like to show its relevance by analyzing
solvable chaotic synchronization systems \cite{umeno1997method,umeno1998superposition}.
Here, a solvable chaotic synchronizing system is such that an invariant
measure, the conditional Lyapunov exponent and the threshold between
synchronization and desynchronization are {\it analytically} obtained for the
unidirectional coupling system.

\subsection{Definition of our system}
As for the first claim, we need three steps to show it.
Firstly, we analytically calculate the
conditional Lyapunov exponent $\lambda$ in our solvable chaotic dynamical system in
accordance with the definition \eqref{eq:pastdif}.
Secondly, we analyze the auxiliary dynamical system similar to
Eq. \eqref{eq:Psi}, and get the exponent $\tilde{\lambda}$ accordance
with Eqs. \eqref{eq:Psi} - \eqref{eq:CLE_solution}.
Thirdly, we confirm that the results obtained by the two types of
analytical methods coincide.

We study the following dynamical system
\begin{eqnarray}
 x(t+1)&=&f(x(t))+\varepsilon\zeta(t),\label{eq:system2}
\end{eqnarray}
where $\varepsilon$ is a coupling parameter, and $\varepsilon\zeta(t)$
correspond to an external forcing input $\xi(t)$.
Firstly we prepare the following solvable chaotic dynamical system \cite{shintani2015,umeno2016exact,umeno2016}:
\begin{eqnarray}
 x(t+1)&=&f(x(t))+\varepsilon\zeta(t)\equiv \phi(x(t),\zeta(t)),\notag\\
f(x(t)) &=& g(x(t)) \equiv \frac{1}{2}\left(x(t)-\frac{1}{x(t)}\right),\label{eq:oursystem}\\
\zeta(t+1) &=& g(\zeta(t)),\notag%\frac{1}{2}\left(\zeta(t)-\frac{1}{\zeta(t)}\right),
\end{eqnarray}
where the function $g$ is associated with the double-angle
formula given by 
$\cot 2\theta=\frac{1}{2}\left(\cot\theta-\frac{1}{\cot\theta}\right)$ \cite{umeno1998superposition}.
The mapping associated with $g$ is a chaotic mapping which has the mixing
property \cite{umeno1998superposition}, and its Lyapunov exponent is $\ln 2$.
The invariant measure of the system $x(t+1)=g(x(t))$ is the standard Cauchy distribution as
follows:
\begin{eqnarray*}
\mu({\rm d}x)  = {\rm C}(x;0,1){\rm d}x,
\end{eqnarray*}
where ${\rm C}(x)$ is Cauchy distribution defined as:
\begin{eqnarray*}
{\rm C}(x;c,\gamma)\equiv \frac{\gamma}{\pi\{(x-c)^2+\gamma^2\}},
\end{eqnarray*}
with $c$ being a median and $\gamma$ a scale parameter.
Therefore, the distribution of variable $x$ in the dynamical system $x(t+1)=g(x(t))$
follows ${\rm C}(x;0,1)$ \cite{umeno1998superposition}, and the distribution of external forcing $\xi$ follows
also ${\rm C}(x;0,1)$.
This is our first solvable chaotic dynamical system.
We see that the conditional Lyapunov exponent changes when
the coupling parameter is varied.
We calculate the conditional Lyapunov exponent of this dynamical
system by the definition \eqref{eq:pastdif} in accordance with
the first step.

\subsection{Conventional ergodic theoretical approach}
We express the conditional Lyapunov exponent of the system
\eqref{eq:oursystem} %denoted
as $\lambda_{g}(\varepsilon)$ since
it will turn out that the conditional Lyapunov exponent crucially
depends on $\varepsilon$.
The conditional Lyapunov exponent $\lambda_{g}(\varepsilon)$ is expressed
as:
\begin{eqnarray*}
 \lambda_{g}(\varepsilon)
   &=& \int_{\mathbb{R}_{x}}\int_{\mathbb{R}_{\xi}}{\rm
     P}(x,\xi)\ln\left|\frac{\partial \phi(x,\xi)}{\partial
			   x}\right|{\rm d}\xi{\rm d}x\\
 &=& \int_{\mathbb{R}_{x}}\int_{\mathbb{R}_{\xi}}{\rm
     P}_{x,\xi}(x){\rm P}_{\xi}(\xi){\rm
     d}\xi\ln\left|\frac{1}{2}\left(1+\frac{1}{x^2}\right)\right|{\rm
     d}x\\
&=& \int_{\mathbb{R}_{x}}{\rm P}_{x,\xi}(x)\ln \left|\frac{1}{2}\left(1+\frac{1}{x^2}\right)\right|{\rm d}x\\
&=& \int_{\mathbb{R}_{x}}{\rm P}_{x,\varepsilon}(x)\ln \left|\frac{1}{2}\left(1+\frac{1}{x^2}\right)\right|{\rm d}x,
\end{eqnarray*}
where ${\rm P}_{x,\xi}(x)$ is the invariant distribution for
the variable $x$ in the superposed dynamical system $x(t+1)=g(x(t))+\varepsilon \zeta(t)$
 with a parameter $\varepsilon$. Note that the superposed distribution is expressed as ${\rm P}_{x,\varepsilon}(x)$ since $\zeta$ is given and the external forcing $\xi$ is only dependent on the coupling parameter $\varepsilon$.

Then we firstly calculate the invariant distribution ${\rm P}_{x,\varepsilon}(x)$,
secondly the conditional Lyapunov exponent $\lambda_{g}(\varepsilon)$,
and thirdly the threshold between synchronization and
desynchronization analytically.

We can calculate the invariant distribution ${\rm
P}_{x,\varepsilon}(x)$ by using the following three
properties, the mixing property for the mapping $g$ (Property 1), the
property of preserving the form of Cauchy
distributions in terms of the Perron-Frobenius equation (PF equation) for $g$
(Property 2), and the property of the stable property for L$\acute{\text{e}}$vy
distributions (Property 3).
Then showing these three properties, we explain how these
properties play roles in order to get ${\rm P}_{x,\varepsilon}(x){\rm d}x$.

As to Property 1, it is already known that the mapping $g$ has mixing
property \cite{umeno1998superposition}.

As for Property 2,
this property implies that $g$ is the mapping which changes an input Cauchy distribution
into {\it another} Cauchy distribution with a {\it different} median and scale parameter. 
We consider the PF equation for the
equation $z=g(x)$ where the input variables $x$ follows ${\rm C}(x;c,\gamma)$.
Since the $g$ is a two-to-one mapping, ${\rm P}_{z}(z)$ satisfies the
following PF equation (Probability Preservation Relation):
\begin{equation*}
   {\rm P}_z(z)|{\rm d}z| ={\rm C}(x_1;c,\gamma)|{\rm d}x_{1}|+{\rm C}(x_2;c,\gamma)|{\rm d}x_{2}|
\end{equation*}
where $x_1$ and $x_2$ $(x_1> x_2)$ are the solutions of the quadratic equation $z=g(x)$. They satisfy the following,
\begin{equation*}
  \begin{cases}
   x_1+x_2=2z\\
  x_1x_2=-1.\\
   \end{cases} 
\end{equation*}
With these relations, the probability distribution ${\rm P}_z(z)$ is
obtained, as the following rescaled Cauchy distribution:
\begin{eqnarray*}%\label{eq:c',gamma'}
{\rm P}_z(z)= {\rm C}(z;c',\gamma'),
\end{eqnarray*}
where $\displaystyle c'=\frac{c(\gamma^2+c^2-1)}{2(\gamma^2+c^2)}$ and $\displaystyle
\gamma'=\frac{\gamma(\gamma^2+c^2+1)}{2(\gamma^2+c^2)}$.

As to Property 3, this stable property has already widely been known.
When distributions of two independent variables $X_1$ and $X_2$ obey a
L$\acute{\text{e}}$vy distribution family, the distribution of the variable
$aX_1+bX_2 (a,b \in \mathbb{R})$ also obeys the same family.
These three properties play roles in order to get the invariant
distribution ${\rm P}_{x,\varepsilon}(x)$.

We can prove that variables $x$ and $\varepsilon\zeta$ in the system
\eqref{eq:oursystem} obey Cauchy distributions respectively with three properties. 
In Appendix B, we describe these three properties in more detail.
By taking into account the Properties 2 and 3, the distribution of the variable
$x$ is changed to a Cauchy distribution with a different median and
scale parameter in every iteration if the initial
input follows a Cauchy distribution.
Hence, we get the following self-consistent recurrence equations about a
median $c$ and a scale parameter $\gamma$ per iteration, as
\begin{equation}\label{eq:parameterr}\left\{
\begin{aligned}
 c(t+1)&=\frac{c(t)(\gamma(t)^2+c(t)^2-1)}{2(\gamma(t)^2+c(t)^2)}\\
 \gamma(t+1)&=\frac{\gamma(t)(\gamma(t)^2+c(t)^2+1)}{2(\gamma(t)^2+c(t)^2)}+|\varepsilon|.
\end{aligned}\right.
\end{equation}
We get the following convergence values $c^*$ and $\gamma^*$ as the stable fixed
point of Eq. \eqref{eq:parameterr} for $t \to \infty$:
\begin{eqnarray*}
 c_g^*&=&0\\
 \gamma_g^*&=&|\varepsilon| + \sqrt{\varepsilon^2 + 1}.
\end{eqnarray*}
Hence, we analytically obtain ${\rm P}_{x,\varepsilon}(x)$ as
the fixed point of the recurrence equation Eq. \eqref{eq:parameterr}:
\begin{eqnarray*}
{\rm P}_{x,\varepsilon}(x)={\rm C}(x;0,\gamma_g^*)  %\label{eq:P(x)_sol}.
\end{eqnarray*}
We should emphasize the following.
Although we assume the initial input follows a Cauchy
distribution in the above analysis, we do not need any restriction for a
distribution of initial point for the system \eqref{eq:system1} because
of the mixing property (Property 1).
Therefore, the invariant distribution ${\rm
P}_{x,\varepsilon}(x)$ is always obtained regardless of any initial points.

\begin{figure}[ht]
  \begin{center}
    \includegraphics[width=80mm]{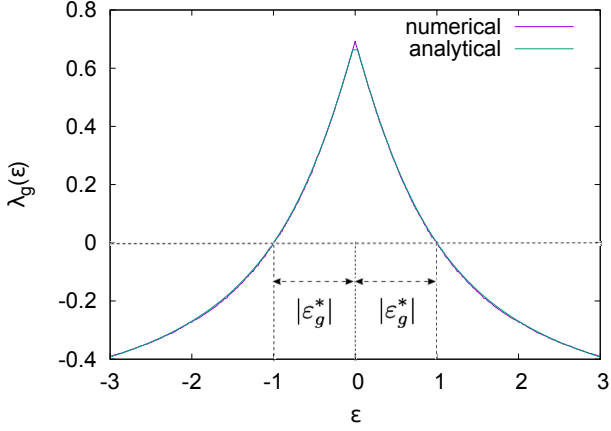}
     \caption{Conditional Lyapunov exponent $\lambda_{g}(\varepsilon)$}
   \label{fig:compare_cle}
  \end{center}
\end{figure}

With ${\rm P}_{x,\varepsilon}(x)$, we can calculate the conditional Lyapunov exponent
$\lambda_{g}(\varepsilon)$ as follows:
\begin{eqnarray*}
\lambda_{g}(\varepsilon) &=& \int_{\mathbb{R}_{x}}{\rm P}_{x,\varepsilon}(x)\ln
 \left|\frac{1}{2}\left(1+\frac{1}{x^2}\right)\right|{\rm d}x\\
 &=& \int_{\mathbb{R}}\frac{\gamma_g^*}{\pi(x^2+{\gamma_g^*}^2)}\ln
  \left|\frac{1}{2}\left(1+\frac{1}{x^2}\right)\right|{\rm d}x\\
  &=& 2\ln\left(\frac{\gamma_g^*+1}{\gamma_g^*}\right)-\ln{2}\\
  &=& 2\ln{(\sqrt{\varepsilon^2+1}-|\varepsilon|+1)}-\ln{2}.
\end{eqnarray*}
Note that $\lambda_g(0)=\ln 2$.
We can also get the threshold $\varepsilon_{g}^*$ between the synchronization and
desynchronization. This is the solution of $\lambda_{g}(\varepsilon_g^*)=0$,
which is
\begin{eqnarray*}
\varepsilon_{g}^*=1.
\end{eqnarray*}
Figure \ref{fig:compare_cle} illustrates that these analytical results coincide with the results of numerical
simulation with the initial condition $x_0=\sqrt{2}$ and $\xi_0=\sqrt{3}$.

\subsection{Proposing ergodic theoretical approach}
As above, we can calculate 
$\lambda_{g}(\varepsilon)$ by Eq. \eqref{eq:pastdif}.
The invariant distribution ${\rm P}(x,\xi)$ is expressed as:
\begin{eqnarray*}
{\rm P}(x,\xi)&=&\frac{\gamma_g^*}{\pi(x^2+{\gamma_g^*}^2)}\frac{|\varepsilon|}{\pi(\xi^2+\varepsilon^2)},
\end{eqnarray*}
because ${\rm P}_{\xi}(\xi)$ is derived from Property 3 for the variable
$\zeta$ that obey ${\rm C}(\zeta;0,1)$.
Then in accordance with
the second step, we would like to calculate $\tilde{\lambda}_{g}(\varepsilon)$
by Eqs. \eqref{eq:Psi} - \eqref{eq:CLE_solution}.

We prepare the following dynamical system by Eq. \eqref{eq:Psi}:
\begin{eqnarray*}%\label{eq:separatesystem}
 \begin{aligned}
  y(t+1)=G_{\xi_0}(y(t))\\
  G_{\xi_0}(y)=g(y)+\xi_0.
\end{aligned}
\end{eqnarray*}
We firstly need to confirm that this system is ergodic for every
$\xi_0$,
which is the sufficient condition for that the 
conditional distribution ${\rm P}_y(y|\xi_0)$ exists for every $\xi_0$.
Hence, we sufficiently calculate three factors, the conditional
distribution ${\rm P}_y(y|\xi_0)$, and the
unique Lyapunov exponents $\lambda_{\Psi}(\xi_0)$, the distribution of external
forcing inputs ${\rm P}_{\xi}(\xi_0)$.

Firstly, we calculate the conditional distribution
${\rm P}_y(y|\xi_0)$.
The invariant measure ${\rm P}_y(y|\xi_0){\rm
d}x$ is also obtained with Properties 1 - 3.

In the same way of getting ${\rm P}_{x,\xi}(x)$, we can prove that ${\rm
P}_y(y|\xi_0)$ also follows a Cauchy distribution.
Hence, we get the following self-consistent
recurrence equations about a median $c$ and a scale parameter $\gamma$, as
\begin{equation*}%\label{eq:parameter''}
 \left\{
\begin{aligned}
 c(t+1)&=\frac{c(t)(\gamma(t)^2+c(t)^2-1)}{2(\gamma(t)^2+c(t)^2)}+\xi_0\\
 \gamma(t+1)&=\frac{\gamma(t)(\gamma(t)^2+c(t)^2+1)}{2(\gamma(t)^2+c(t)^2)}.
\end{aligned}\right.
\end{equation*}
We get the following convergence values $\hat{c}$ and $\hat{\gamma}$ as the stable fixed
point for $t \to \infty$: %They satisfies following:
\begin{equation*}
\begin{aligned}&\left\{
 \begin{aligned}
  \hat{c}&=\xi_0\\
  \hat{\gamma}&=\sqrt{1-\xi_0^2}
 \end{aligned}\ \ \ \ \ \ \  (\text{if}\ |\xi_0|<1),\right. \\
 &\left\{
 \begin{aligned}
  \hat{c}&=\xi_0+\text{sgn}(\xi_0)\sqrt{\xi_0^2-1}\\
  \hat{\gamma}&=0
 \end{aligned}\  (\text{if}\  |\xi_0|\ge 1),\right.
 \end{aligned}
\end{equation*}
and $\text{sgn}(\xi_0)=\left\{\begin{matrix}
			     1\ \ \text{if} \ \xi_0>0.\\
			     -1\ \text{if}\  \xi_0<0.
			     \end{matrix} \right.$\\
Thus, we obtain the conditional
distribution ${\rm P}(y|\xi_0)$ as follows:
\begin{equation*}%\label{eq:P(x|xi)}
  {\rm P}_y(y|\xi_0)=\left\{
  \begin{aligned}
   &{\rm C}(y;\xi_0,\sqrt{1-\xi_0^2}) &(|\xi_0|<1).\\
   & {\rm C}(y;\xi_0+{\rm sgn}(\xi_0)\sqrt{\xi_0^2-1}),0)  &(|\xi_0|\ge1).
    \end{aligned}\right.
\end{equation*}

As for unique Lyapunov exponents $\lambda_{\Psi}(\xi_0)$,
we can calculate them as follows:
\begin{eqnarray*}
 \lambda_{G}(\xi_0) &=& \int_{\mathbb{R}_{y}}{\rm P}_y(y|\xi_0)\ln{\left|\frac{1}{2}\left(1+\frac{1}{y^2}\right)\right|}{\rm
 d}y\notag\\
 & =& \begin{cases}
    \ln(1+\sqrt{1-\xi_0^2}) \ \ \ \ (|\xi_0|<1)\\
       \ln\left\{1+\frac{1}{(\xi_0+{\rm
     sgn}(\xi_0)\sqrt{\xi_0^2-1})^2}\right\}-\ln 2 \ (|\xi_0|\ge1).
 \end{cases}%\label{eq:lambda_\Psi}
\end{eqnarray*}

\begin{figure}[ht]
  \begin{center}
    \includegraphics[width=80mm]{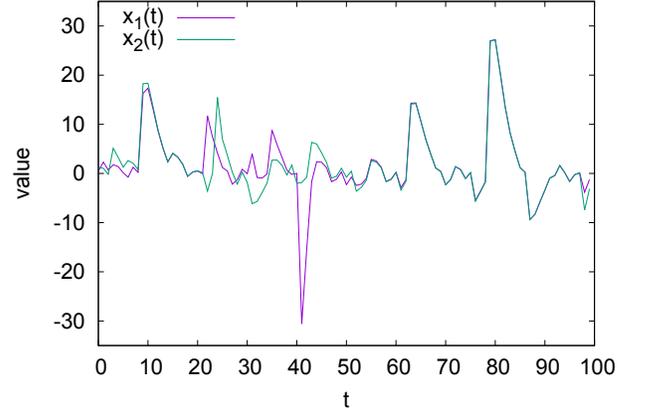}
    \begin{center}
     \caption{Partial synchronization in the system \eqref{eq:oursystem} with $\varepsilon=0.8$}
      \label{fig:synchronize}
    \end{center}
  \end{center}
\end{figure}

\begin{figure}[ht]
  \begin{center}
    \includegraphics[width=80mm]{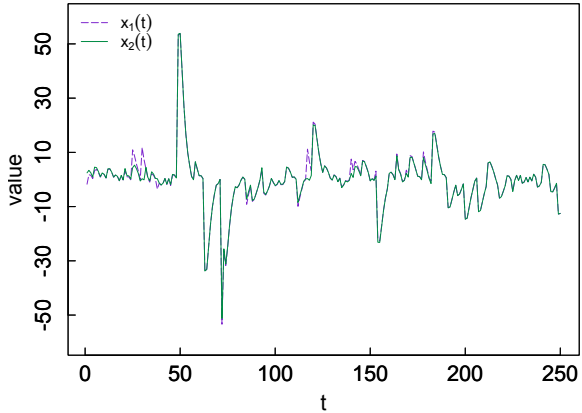}
    \begin{center}
     \caption{Chaotic synchronization in the system \eqref{eq:oursystem} with $\varepsilon=1.05$ ($x_1(0)=-\sqrt{2},x_2(0)=\sqrt{5}$)}
      \label{fig:synchronize2}
    \end{center}
  \end{center}
\end{figure}

Thus, we can calculate the exponent $\tilde{\lambda}_g(\varepsilon)$ according to Eq. \eqref{eq:CLE_solution}:
\begin{eqnarray*}
 \tilde{\lambda}_g(\varepsilon) &=& \int_{\mathbb{R}_{\xi}}{\rm P}_{\xi}(\xi_0)\lambda_{G}(\xi_0){\rm
  d}\xi_0\\
&=&  \int_{\mathbb{R}}\frac{|\varepsilon|}{\pi(\xi_0^2+\varepsilon^2)}\lambda_{G}(\xi_0){\rm
  d}\xi_0\\
 &=& 2\ln{(\sqrt{\varepsilon^2+1}-|\varepsilon|+1)}-\ln{2}\\
 &=& \lambda_{g}(\varepsilon).
\end{eqnarray*}
As above, we confirm that our first main claim is certainly satisfied.
This claim signifies that a conditional Lyapunov exponent is expressed as
the ensemble average of the set of unique Lyapunov exponents
of the auxiliary dynamical system.
Figure \ref{fig:synchronize} illustrates that two different trajectories with
different initial points ($x_1(0)=0.46,x_2(0)=1.3$)  partially synchronize.
The coupling parameter $\varepsilon$ is $0.8$.
Note that although the conditional Lyapunov exponent is positive in FIG. \ref{fig:synchronize}, the partial synchronization occurs since it is the averaged factor of unique Lyapunov exponents.
When the conditional Lyapunov exponent is negative, the infinitesimal synchronization error converges to $0$ for $t\to\infty$ (see FIG. \ref{fig:synchronize2}).

\section{Example2 for the first claim}
Our analysis is not restricted to the above example.
We consider
another solvable chaotic dynamical system, and can also confirm
that the conditional Lyapunov exponent in the system is obtained by
Eq. \eqref{eq:conclusion1}.
This dynamical system is defined as follows:
\begin{eqnarray*}
 x(t+1)&=&f(x(t))+\varepsilon\zeta(t),\notag\\
f(x(t)) &=& h(x(t)) \equiv \frac{2x(t)}{1-x(t)^2} ,\\ %\label{eq:ftan}  \\ 
 \zeta(t+1) &=& h(\zeta(t)),\notag\\
     H_{\xi_0}(y)&=&h(y)+\xi_0.
\end{eqnarray*}
The chaotic mapping $h$ is associated with the double formula
$\tan 2 \theta=\frac{2\tan\theta}{1-\tan^2\theta}$ \cite{umeno1998superposition}. Its Lyapunov
exponent is $\ln 2$, and the invariant measure
obeys the standard Cauchy distribution.
Then, the invariant distribution ${\rm P}_{x,\xi}(x)$ and the conditional Lyapunov
exponent $\lambda_{h}(\varepsilon)$, and the threshold $\varepsilon_{h}^*$
are given as:
\begin{eqnarray*}
 {\rm P}_{x,\xi}(x)&=&{\rm C}(x;0,\gamma_{h}^*),\\
 \lambda_{h}(\varepsilon)&=&2\ln\left(\frac{\gamma_{h}^*+1}{{\gamma_{h}^*}^2+1}\right)+\ln{2},\\
 \varepsilon_{h}^*&=&0.78\cdots,
\end{eqnarray*}
where $\gamma_{h}^*$ is the positive number which satisfies the following
equation:
\begin{eqnarray*}
{\gamma_{h}^*}^3-|\varepsilon|{\gamma_{h}^*}^2-\gamma_{h}^*-|\varepsilon|=0.
\end{eqnarray*}

\section{Example for the second claim}
Here, as for the second claim,
the conditional Lyapunov exponent for a system \eqref{eq:system1} is characterized by only
two factors, a dynamical system associated with $f$ and a distribution of external
forcing input $\xi$.

In order to show this claim we consider
six different dynamical systems. Each of them has a
different chaotic mapping $f$ or different system for $\xi$ as follows:
\begin{eqnarray*}
 f(x)=
  \left\{
  \begin{aligned}
   &g(x)=\frac{1}{2}\left(x-\frac{1}{x}\right),\\
   &h(x)=\frac{2x}{1-x^2},
\end{aligned}\right.
\end{eqnarray*}
\begin{eqnarray*}
  \left\{
   \begin{matrix}
\displaystyle \zeta(t+1) = g(\zeta(t)),\\
\displaystyle  \zeta(t+1) = h(\zeta(t)),\\
\displaystyle    \bm{\zeta} = \text{Crand}(0,1),
    \end{matrix}\right.
 \end{eqnarray*}
 where Crand$(c,\gamma)$ is a set of random numbers which follow ${\rm
 C}(\zeta;c,\gamma)$.
 The algorithm to get Crand$(c,\gamma)$ follows:
 \begin{eqnarray*}
 \text{Crand}(0,1) = \tan\left(\frac{\pi}{2}({\rm U}(-1:1))\right),
 \end{eqnarray*}
 where ${\rm U}(-1:1)$ are uniform random numbers on the interval $(-1:1)$.

\begin{figure}[htb]
   \begin{center}
    \includegraphics[width=85mm]{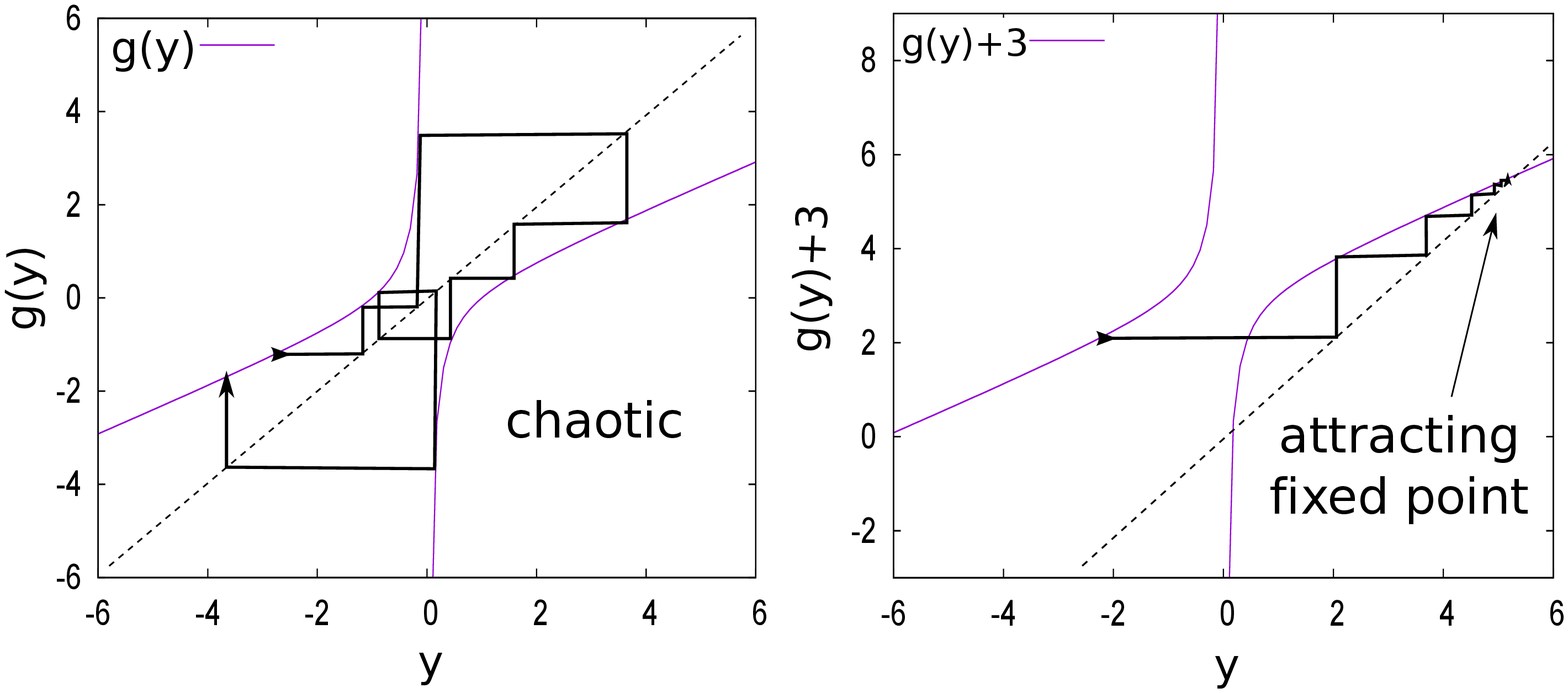}
     \caption{Attracting fixed point is generated by changing $\xi_0$ in
    $G_{\xi_0}(y)$}
       \label{fig:f1_attractor}
  \end{center}
   \begin{center}
	  \includegraphics[width=85mm]{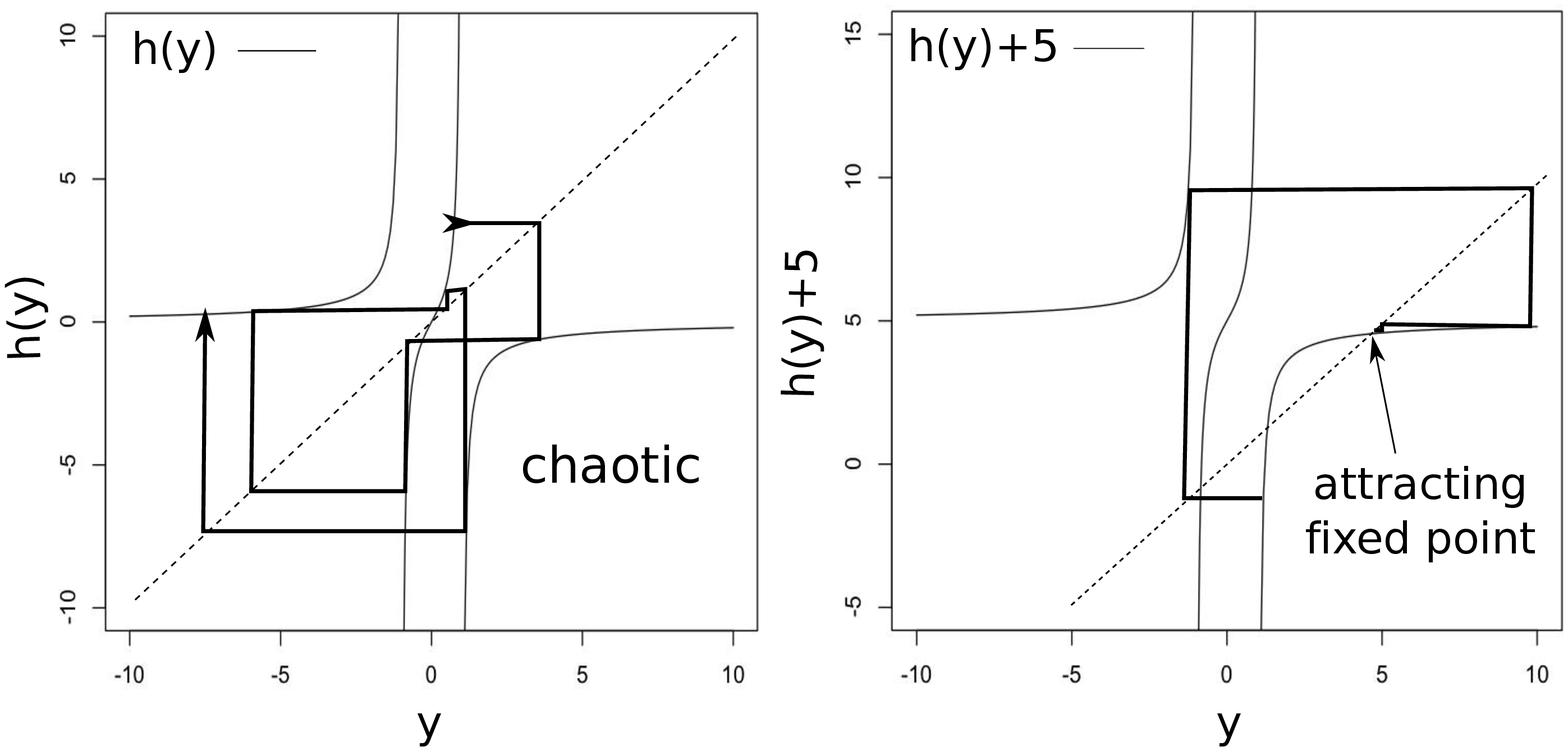}
	  \caption{Attracting fixed point is generated by changing
    $\xi_0$ in $H_{\xi_0}(y)$}
   \label{fig:f2_attractor}
   \end{center}
\end{figure}

\begin{figure}[htb]
   \begin{center}
        \includegraphics[width=85mm]{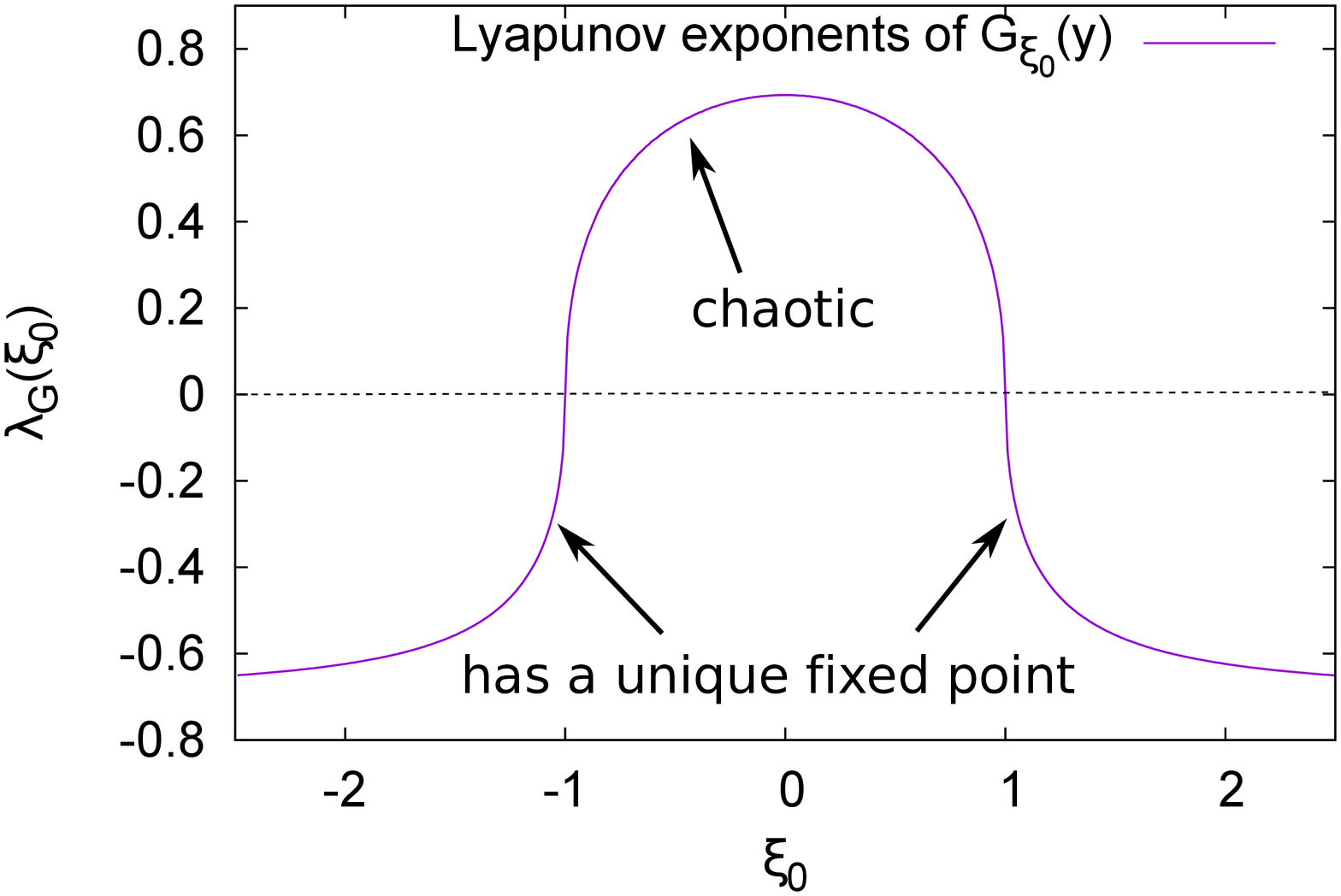}
    \caption{Unique Lyapunov exponents of $G_{\xi_0}(y)$}
    \label{fig:g_Lyapunov}
   \end{center}
   \begin{center}
    	  \includegraphics[width=90mm]{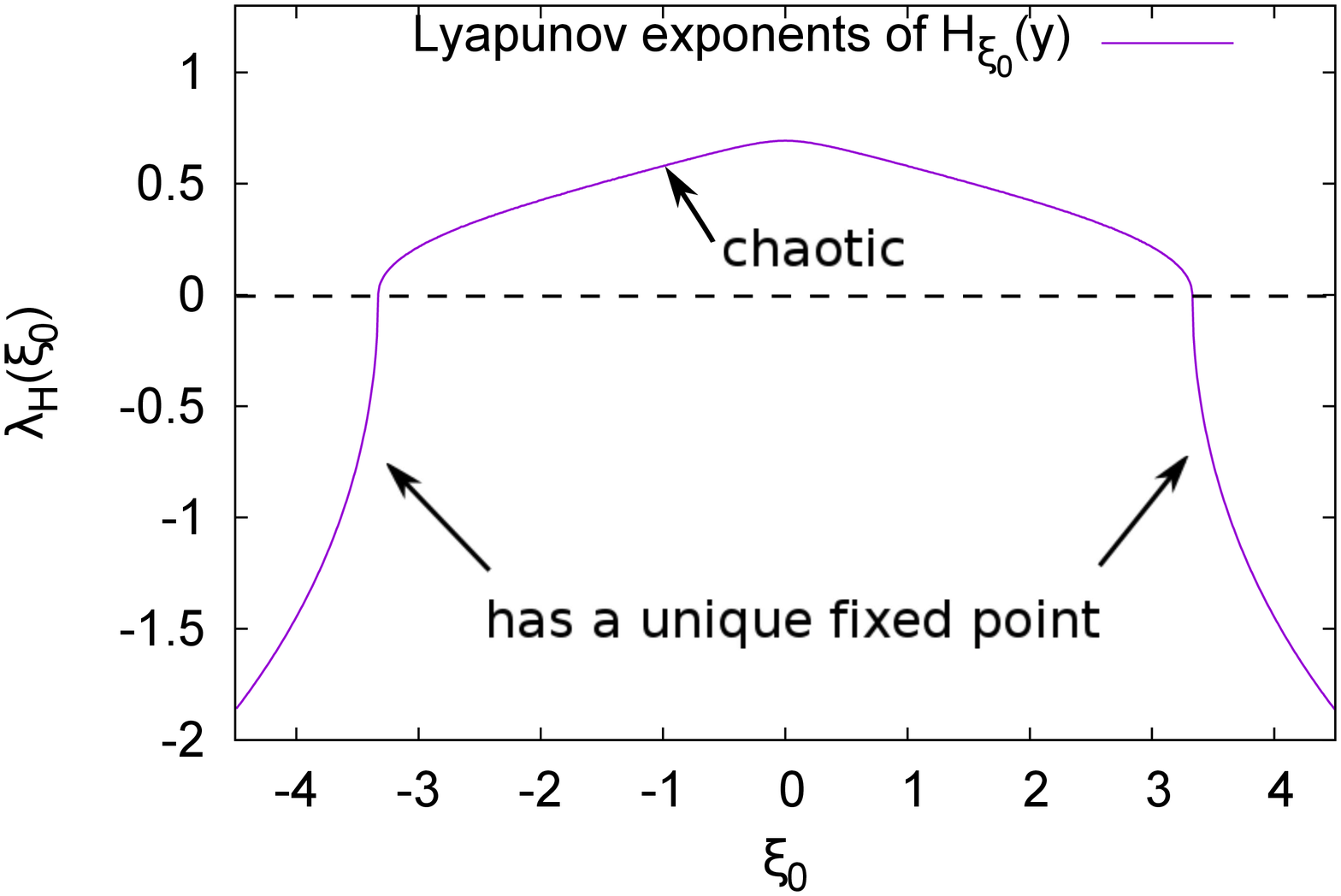}
    \caption{Unique Lyapunov exponents of $H_{\xi_0}(y)$}
    \label{fig:h_Lyapunov}
   \end{center}
\end{figure}

 We compare the conditional Lyapunov exponent of these systems with
 Table \ref{tab:relation}.
\begin{table*}
    \begin{center}
\caption{Relations among the combination of $f$, $\xi$ and
     the conditional Lyapunov exponents}
  \begin{tabular}{|p{7mm}|p{18mm}||p{15mm}|p{30mm}|p{18mm}|p{18mm}|p{32mm}|p{10mm}|}
  \hline
  $f$ & Generation Mechanism of $\zeta$ & Lyapunov exponent of $f$ & The invariant distribution of the system
      $x(t+1)=f(x(t))$ & ${\rm P}_{\xi}(\xi)$ &
	      ${\rm P}_{x,\xi}(x)$ & the property of time series in external
			  forcing input $\xi$ &
		      $\lambda$\\ \hline \hline
 $g$ & $g$ & $\ln 2$ & ${\rm C}(x;0,1)$ & ${\rm C}(\xi;0,|\varepsilon|)$ &
	      ${\rm C}(x;0,\gamma_g^*)$ & chaotic & $\lambda_{g}(\varepsilon)$\\ \hline
 $g$ & $h$ & $\ln 2$ &${\rm C}(x;0,1)$ & ${\rm C}(\xi;0,|\varepsilon|)$ &
	  ${\rm C}(x;0,\gamma_g^*)$  & chaotic & $\lambda_{g}(\varepsilon)$ \\
  \hline
  $g$ & Crand(0,1) & $\ln 2$ &${\rm C}(x;0,1)$ & ${\rm C}(\xi;0,|\varepsilon|)$ &
	  ${\rm C}(x;0,\gamma_g^*)$  & random & $\lambda_{g}(\varepsilon)$ \\ \hline
  $h$ & $g$ & $\ln 2$ & ${\rm C}(x;0,1)$ & ${\rm C}(\xi;0,|\varepsilon|)$ &
	  ${\rm C}(x;0,\gamma_{h}^*)$  & chaotic  & $\lambda_{h}(\varepsilon)$ \\ \hline
  $h$ & $h$ & $\ln 2$ &${\rm C}(x;0,1)$ & ${\rm C}(\xi;0,|\varepsilon|)$ &
	  ${\rm C}(x;0,\gamma_{h}^*)$  & chaotic & $\lambda_{h}(\varepsilon)$ \\ \hline
  $h$ & Crand(0,1) & $\ln 2$ & ${\rm C}(x;0,1)$ & ${\rm C}(\xi;0,|\varepsilon|)$ &
	  ${\rm C}(x;0,\gamma_{h}^*)$  & random &
			      $\lambda_{h}(\varepsilon)$ \\ \hline
  \end{tabular}
     \label{tab:relation}
    \end{center}
\end{table*}
 As we can see from the Table \ref{tab:relation},
we can find that some factors do not influence the
 conditional Lyapunov exponent such as the property of time series in
 external forcing input.
 Then, we show that the conditional Lyapunov exponent is
 characterized by the original mapping $f$ and the distribution of
 external forcing input $\xi$.
The original dynamical system is changed to the system which has an
attracting fixed point because of the external forcing input.
The conditional Lyapunov exponent is changed by the existence of attracting fixed points.
Figures \ref{fig:f1_attractor} and \ref{fig:f2_attractor} illustrate that
the mappings $G$ and $H$ are changed to ones that have
attracting fixed points by external forcing input  respectively.
Attractors of systems depend on how attracting fixed points are generated.
The created attractor leads to the $\xi$-dependence of
$\lambda_\Psi(\xi_0)$.
Figures \ref{fig:g_Lyapunov} and \ref{fig:h_Lyapunov} show the set of Lyapunov exponents $\lambda_{G}(\xi_0)$
and that of $\lambda_{H}(\xi_0)$, respectively.
Hence, the original mappings determine how attracting fixed points are generated. 
Furthermore, with Eq. \eqref{eq:conclusion1} from the first claim, the conditional
Lyapunov exponent is calculated by the set of Lyapunov exponents
$\{\lambda_\Psi(\xi_0)|\xi_0\in \mathbb{R}_{\xi}\}$ and the
distribution of the external forcing input ${\rm P}_\xi$.
As above, the conditional Lyapunov exponent in system \eqref{eq:system1}
is uniquely characterized by only two factors, the original dynamical
system and the distribution of external forcing input.

 \section{Conclusion}
 We have two main claims. First, the conditional Lyapunov exponent is
 expressed as Eq. \eqref{eq:conclusion1} provided the ergodicity in
 dynamical systems. This yields that the conditional Lyapunov exponent is expressed as
 the ensemble average of the set of unique Lyapunov exponents of the
 auxiliary dynamical system \eqref{eq:Psi}.
 Second, the conditional Lyapunov exponent is characterized only two factors, an original dynamical
 system and a distribution of external forcing input.
 Then, although we consider CS in one-dimensional unidirectionally
coupled  dynamical systems only, this claim
 will also hold in multi-dimensional chaotic systems in terms of ergodic theory.

\begin{acknowledgments}
The authors thank Shin-itiro Goto (Kyoto University) for stimulating
 discussions. 
\end{acknowledgments}

\appendix

\section{Ergodic dynamical systems}
In this Appendix a short introduction of ergodic theory is provided. 
Since this appendix aims at providing clear definitions of terminologies used 
in the main text,   
this appendix is not comprehensive.  
For a comprehensive review from a mathematical viewpoint (see 
Refs.\cite{arnold-avez,boyarsky}), 
also from a physical viewpoint (see Ref.\cite{eckmann}).
 
%%%%%%%%%%%%%%%%%%%%%%%%%%%%%%%%%
%\subsection{Ergodic dynamical systems }
%%%%%%%%%%%%%%%%%%%%%%%%%%%%%%%%%%5
\begin{Def}
%%%%%%%%%%
(Dynamical system, \cite{arnold-avez}):
A dynamical system $(\cM,\mu,\varphi_{\,t})$ is a measure-space $(\cM,\mu)$ 
equipped with a one-parameter group $\varphi_{\,t}$ of automorphisms 
(except for spaces of measure-zero) 
of $(\cM,\mu)$, $\varphi_{\,t}$ depending measurably
of $t$. Here the parameter $t$ denotes an integer.

% Let $(\cM,\mu)$ be a measure space with $\mu$ being a measure, $\varphi_{\,t}$ 
% a one-parameter group with $t\in\mbbR$ 
% that preserves $\mu$ except for spaces for measure zero
% Then $(\cM,\mu,\varphi_{t})$ is called a dynamical system.
%%%%%%%%%%
\end{Def}
%%%%%%%%%
Given a dynamical system $(\cM,\mu,\varphi_{\,t})$, it follows that  
$\mu(\varphi_{\,t}A)=\mu(A)$, where $A$ is a measurable set, and that 
$\varphi_{\,t}$ is a measurable in $\cM\times\mbbR$. 
In what follows $\mu(\cM)=1$ is assumed. 

The following average often appears in physics. 
%%%%%%%%%%
\begin{Def}
%%%%%%%%%%
(Time-average, \cite{arnold-avez}): 
Let $(\cM,\mu,\varphi_{t})$ be a dynamical system, and $f$ 
a complex-valued function defined on $\cM$. If there exists the quantity 
$$
f^{\,*}(x)
=\lim_{N\to\infty}\frac{1}{N}\sum_{n=0}^{N-1}f(\varphi^{\,n}x),\qquad 
x\in\cM,\quad n\in\mbbZ
$$ 
then $f^{\,*}(x)$ is called the time-average of $f$.
%%%%%%%%%
\end{Def}
%%%%%%%%%
Also the following average often appears and is related 
to the time-average for some dynamical systems. 
%%%%%%%%%%%
\begin{Def}
\label{definition-space-average}
%%%%%%%%%%%
(Space-average, \cite{arnold-avez}):  
Let $(\cM,\mu,\varphi_{t})$ be a dynamical system, and $f$ 
a complex-valued function defined on $\cM$. If there exists the quantity 
$$
\overline{f}
=\int_{\cM}f(x)\,\mu(\dr x)\qquad 
x\in\cM,\quad
$$ 
then $\overline{f}$ is called the space-average of $f$.
%%%%%%%%%%
\end{Def}
%%%%%%%%%%
Space-average defined above is called ensemble average in the main text. 
The following is used in the main text.
%%%%%%%%%%%%%
\begin{Def}
%%%%%%%%%%
(Absolutely continuous function with respect to the Lebesgue measure): 
In Definition\, \ref{definition-space-average}, 
if $\mu(\dr x)$ is of the form 
$$
\mu(\dr x)
=\rho(x)\,\dr x,
$$ 
then $\rho$ is called an absolutely continuous function with respect to the Lebesgue 
measure $\dr x$.
%%%%%%%%%%%
\end{Def}
%%%%%%%%%%

We are now ready to state the definition of ergodic system.
%%%%%%%%%%%%%%
\begin{Def}
%%%%%%%%%%%%
(Ergodic system, \cite{arnold-avez}): 
Let $(\cM,\mu,\varphi_{\,t})$ be a dynamical system. 
If the following condition is satisfied
$$
f^{\,*}(x)
=\overline{f}, 
$$
for  any integrable function  $f$ in the sense of $f\in L_{\,1}(\cM,\mu)$, 
then the dynamical system is 
called an ergodic system.
%%%%%%%%%
\end{Def}
%%%%%%%%%
This states that for an ergodic dynamical system, one can replace the 
time-average of $f$ with the space-average of it.
An example of how to apply this property is Eq. \eqref{eq:pastdif} in the main text. 

The following property is a stronger property for dynamical systems.
%%%%%%%%%%%
\begin{Def}
%%%%%%%%%%%%%5
(Mixing system, \cite{arnold-avez}):  
Let $(\cM,\mu,\varphi_{\,t})$ be a dynamical system. 
If the condition 
$$
\lim_{t\to\infty }\mu\left[\,
\varphi_{\,t}A\cap B
\,\right]
=\mu(A)\,\mu(B),
$$
is satisfied for all measurable set $A$ and $B$,  
then the dynamical system is called a mixing system.\\
%%%%%%%%%%
\end{Def} 
%%%%%%%%%%

\section{Three Properties}
In this Appendix, we 
describe the Properties 2 and 3 in Section III in the main text in more detail.
About the Property 2, we get ${\rm P}_z(z)$ in Section III-A as follows:
\begin{eqnarray*}%\label{eq:c',gamma'}
{\rm P}_z(z)&=& \frac{\gamma}{\pi((x_1-c)^2+\gamma^2)}\left|\frac{1}{\frac{{\rm d}z}{{\rm d}x_1}}\right|+\frac{\gamma}{\pi((x_2-c)^2+\gamma^2)}\left|\frac{1}{\frac{{\rm d}z}{{\rm d}x_1}}\right|\\
 &=&
{\rm C}(x_1;c,\gamma)\frac{2x_1^2}{x_1^2+1}+{\rm C}(x_2;c,\gamma)\frac{2x_2^2}{x_2^2+1}
   \\
 &=&
 \frac{\frac{\gamma(\gamma^2+c^2+1)}{2(\gamma^2+c^2)}}{\pi\left\{\left(z-\frac{c(\gamma^2+c^2-1)}{2(\gamma^2+c^2)}\right)^2+\left(\frac{\gamma(\gamma^2+c^2+1)}{2(\gamma^2+c^2)}\right)^2\right\}}
 \\
&=& {\rm C}(z;c',\gamma').
% &=& \\
\end{eqnarray*}
This shows that the mapping $g$ changes the median $c$ and the scale parameter $\gamma$ of the input Cauchy distribution as:
\begin{eqnarray*}
 &\text{median}&:\  c\  \to\  \frac{c(\gamma^2+c^2-1)}{2(\gamma^2+c^2)}
  \ (\equiv c')\\
 &\text{scale parameter}&:\  \gamma\  \to\
  \frac{\gamma(\gamma^2+c^2+1)}{2(\gamma^2+c^2)} \ (\equiv \gamma').
\end{eqnarray*}

We utilize the Properties 1-3 to get Eq. \eqref{eq:parameterr}.
When the variables $x(t)$ which follow a Cauchy distribution $\text{C}(c(t),\gamma(t))$ in the 
dynamical system Eq. \eqref{eq:system2}, the variables $x(t+1)$ which also follow a Cauchy distribution as Fig. \ref{fig:transition1}.

Hence, we get the self-consistent recurrence equations Eq. \eqref{eq:parameterr} about a
median $c$ and a scale parameter $\gamma$ per iteration.
This idea is also utilized to get ${\rm P}_y(y|\xi_0)$ in Section III-C.
\begin{figure*}[h]
    \begin{center}
     \begin{eqnarray*}
      \begin{array}{ccccc}
 &&\underbrace{x(t)}_{{\rm C}(c(t),\gamma(t))}&&\underbrace{\zeta(t)}_{{\rm C}(0,1)}\\
 && \downarrow&&\\
 &&\underbrace{f(x(t))}_{{\rm C}(c'(t),\gamma'(t)) \left(\raisebox{1.2ex}{.}\raisebox{.2ex}{.}\raisebox{1.2ex}{.}\  \text{Property
  2}\right)}&&\underbrace{\varepsilon\zeta(t)}_{{\rm C}(0,|\varepsilon|) \left(\raisebox{1.2ex}{.}\raisebox{.2ex}{.}\raisebox{1.2ex}{.}\ \text{Property 3}\right)}\\
  && \downarrow&&\\
\underbrace{x(t+1)}_{{\rm C}(c(t+1),\gamma(t+1))
  \left(\raisebox{1.2ex}{.}\raisebox{.2ex}{.}\raisebox{1.2ex}{.}\
   \text{Property
   3}\right)}&=&\underbrace{f(x(t))}_{{\rm C}(c'(t),\gamma'(t))}&+&\underbrace{\varepsilon\zeta(t)}_{{\rm C}(0,|\varepsilon|)}
  \end{array}
     \end{eqnarray*}
     \caption{Propagation of Cauchy distributions from $x(t)$ and
     $\zeta(t)$ to $x(t+1)$}
     \label{fig:transition1}
    \end{center}
\end{figure*}
%}
%%%%%%%%%%%%%%%%%%%%%%%%%%%

%%%%%%%%%%%%%%%%%%%%%
\end{document}